# Unveiling the Influence of Graphitic Nanocarbon Structures, Defects, and Surface Functional Groups on Supercapacitive Performance in LiTFSI Electrolyte

Phongphot Sakulaue[a,b,f], Bénédicte Réty[b,c,d], Takafumi Ishii[e], Camelia Matei Ghimbeu[b,c,d*], Khanin Nueangnoraj[a*]

[a] *School of Bio-Chemical Engineering and Technology, Sirindhorn International Institute of Technology, Thammasat University, Pathum Thani, 12120, Thailand.*

[b] *Université de Haute-Alsace, Institut de Science des Matériaux de Mulhouse (IS2M), CNRS UMR 7361, F-68100 Mulhouse, France.*

[c] *Université de Strasbourg, F-67081 Strasbourg, France.*

[d] *Réseau sur le Stockage Electrochimique de l'Energie (RS2E), CNRS FR3459, 80039 Amiens Cedex, France*

[e] *International Research and Education Center for Element Science, Faculty of Science and Technology, Gunma University, Japan.*

[f] *Division of Chemical Engineering, Faculty of Engineering, Rajamangala University of Technology Krungthep, Bangkok 10120, Thailand.*

*Corresponding authors: C.M. Ghimbeu (camelia.ghimbeu@uha.fr) and K. Nueangnoraj (khanin@siit.tu.ac.th)

**TOC Graphic**

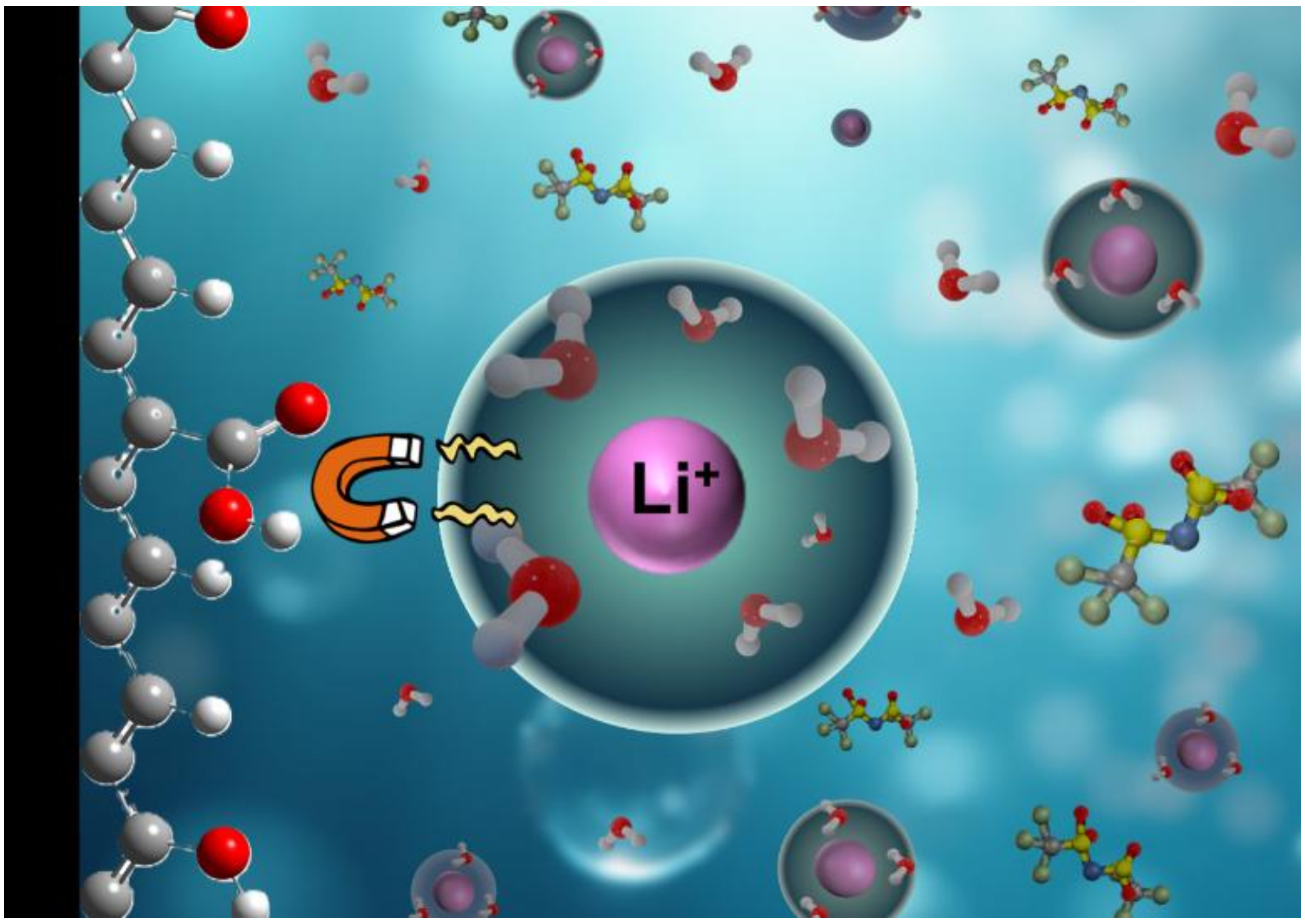

Uncover the previously overlooked impact of carbon defects and oxygen functionalities on supercapacitor performance in LiTFSI electrolytes.

**Abstract**

When investigating the performance of supercapacitors, it was found that they offer high power density and long life but are limited in capacity and energy density. Water-in-salt electrolytes, i.e., 21 m lithium bis(trifluoromethanesulfonyl) imide (LiTFSI), offer a solution by expanding the electrochemical window compared to a diluted aqueous environment. While significant research has been dedicated to improving these electrolytes, the impact of carbon properties on energy storage remains relatively underexplored. This study investigated the influence of carbon structures, defects, and surface functional groups on supercapacitive performance in LiTFSI electrolytes. For this purpose, a series of carbon materials, including carbon nano-onions (CNOs) and porous conductive carbon black (ENSACO), were synthesized and systematically modified by oxidative and reductive thermal treatments. Our results indicated that higher specific surface area and active surface area improve the ion adsorption and capacitance, while specific oxygen functionalities, such as carboxylic and anhydride groups, were found to significantly enhance capacitance. Furthermore, crystallinity and graphitic characteristics play a role in capacitance, with smaller graphene domains and greater disorder promoting ion storage. The study demonstrates a direct correlation between carbon microstructure and electrochemical behavior, providing insights into the rational design of carbon electrodes in high-performance supercapacitors.

## 1. Introduction

Electric double-layer capacitors (EDLCs), also known as supercapacitors (SCs), are emerging as promising energy storage devices for high power rates in our energy-driven lives (1-4). They consist of a positive and a negative electrode separated by a separator, a membrane that is electronically not conductive but highly permeable for ion transport (5-7). The stored energy is the result of the electric field generated by opposing charges that are created on the electrode surface, when the charging voltage is applied. This charge accumulation is favored by the presence of an electrolyte that allows the ions to move between the electrodes, enabling the storage and release of electrical energy (8, 9). SCs can generally offer high power density and fast charging, but have limited capacity, low energy density and high energy decay (10-13). Recent research efforts have primarily focused on improving the energy density of supercapacitors without sacrificing their high-power density. The energy density (Wh $kg^{-1}$) is influenced by the specific capacitance ($C_s$, in F $g^{-1}$), and the square of the cell voltage ($\Delta V^2$). Since voltage is a squared parameter, even small increases can lead to significant improvements in energy density. While the capacitive performance depends on the ability of the electrode to store charge ions, the cell voltage is limited by the electrochemical properties of the electrolyte (14). In aqueous electrolytes, for example, the electrochemical stability window of water is limited to 1.23 V, which restricts the overall voltage (15).

To overcome these limitations, water-in-salt electrolytes (WiSEs) have been used, especially those containing lithium bis(trifluoromethanesulfonyl) imide (LiTFSI) at a high concentration of 21 m. The reduction of the $H_2O/Li^+$ molar ratio in these highly concentrated electrolytes has a strong influence on the solvation structure of the electrolyte, leading to the suppression of both the hydrogen evolution reaction (HER) and the oxygen evolution reaction (OER) (16-20). This suppression extends the operating voltage window of the devices, as recent research has shown (21-25). For example, supercapacitors with 21 m LiTFSI have shown stable operation at voltages of more than 2 V with gravimetric capacitances of over 100 F $g^{-1}$ (26). In addition, these electrolytes have enabled energy densities of approximately 60.6 W h $kg^{-1}$ at room temperature, increasing to 78.9 W h $kg^{-1}$ at 120 °C (27), significantly outperforming traditional aqueous electrolytes. Recent studies have shown that supercapacitors using highly concentrated LiTFSI-based WiSEs exhibit long-term cycling stability, with devices employing 21 m LiTFSI enduring up to 10,000 charge–discharge cycles while retaining more than 80% of their original capacitance (27, 28).

Although several research efforts have focused on improving the performance of SCs using LiTFSI-based WiSEs, there is still a gap in our understanding of the intricate mechanisms that drive capacitive energy storage towards the carbon structure electrode. In particular, the influence of carbon textures and structures, including factors such as surface area, pore size distribution, graphitic domains and surface chemistry, has not yet been thoroughly explored. Therefore, the unique characteristics of LiTFSI-based WiSEs, including their high concentration and potential operating voltage window, require an in-depth investigation of the influence of carbon electrodes and capacitance behavior. Surface area has been widely recognized as an important factor in determining carbon capacitance. It is well known that capacitance increases with a larger carbon surface area (14, 29-31). In addition, pore size distribution also plays a crucial role in improving capacitance performance. The distribution of micropores allows the carbon structure to approach that of bare electrolyte ions, thereby increasing capacitance (32-34). In addition, it has been observed that the presence of mesopores facilitates the transport of electrolyte ions by serving as an electrolyte reservoir, thus improving access to microporosity and increasing rate capability (30).

On the other hand, Moussa *et al*. investigated the influence of carbon surface chemistry and defects on capacitance in aqueous and organic electrolytes. The carbon defects in this case are dislocations, atomic vacancies or stacking faults, which are mainly located in the edge planes of the carbon and form the so-called active sites or active surface area (ASA) (29, 35, 36). The use of carbon nano-onions (CNOs) as model carbon was due to their fully accessible surface for ion adsorption, as there is no porous network inside the particles. This made it possible to precisely tune the active surface and surface functionality of CNOs to investigate the relationship between structural defects and surface functionality over capacitance. Interestingly, the presence of ASA can improve the gravimetric capacitance in both aqueous and organic electrolytes. However, the surface functional groups can improve the capacitance in an aqueous electrolyte, but have less influence in an organic electrolyte (29). In addition, Tang *et al*. suggested that the origin of carbon corrosion in organic electrolytes is H-terminated edge sites (such as phenol, ether, and carbonyl) at the positive electrode, while the initial degradation at the negative electrode occurs mainly at the basal planes of carbon rather than the edge sites (37, 38).

Understanding the abovementioned relationships is therefore essential for the rational design and optimization of carbon electrode materials, especially for high-performance SCs with LiTFSI electrolyte. The aim of this research is to investigate the effects of carbon properties on the capacitance in LiTFSI electrolytes. For this purpose, two structurally different carbon materials were selected: CNOs, which are nanometric and graphitic materials, and carbon black (ENSACO), a classic porous and disordered carbon. These materials represent a broad spectrum of carbon architectures and allow to comprehensively investigate the relationships between structural features and electrochemical performance. By systematically investigating the influence of carbon textures, structures, defects, and surface functional groups on the capacitance behavior in the LiTFSI electrolyte, our study not only provides a new insight into the essential properties of carbon for the LiTFSI electrolyte but also serves as a guide to understand the electrochemical behavior in other WiSE systems.

## 2. Experimental

### *2.1 Carbons synthesis and modification*

A family of several carbons, derived from two types of carbon materials i.e., carbon nano-onions (CNOs) and conductive carbon black (ENSACO350, IMERYS) were used in this study. CNOs were synthesized using ultrapure nanodiamond particles (NDs, ADAMAS Nanotechnologies Inc.) with a purity of ≥ 98%. NDs were heated under argon gas at 1,500 °C for 1 hour with a flow rate of 10 L $h^{-1}$, to synthesize CNOs, resulting in the sample designated as CNOs-1500. To modify the surface chemistry, the CNOs thus produced were thermally treated at 460 °C in air for 1 hour. During the heating and cooling stages, an argon atmosphere (flow rate 10 L $h^{-1}$) was maintained to avoid unwanted oxidation. The sample obtained from this process was denoted as CNOs-1500-Air. A fraction of the sample was then annealed with $H_2$ at 900 °C for 1 hour to remove the oxygen functional groups and was labeled as CNOs-1500-Air-$H_2$.

A similar modification was carried out with the ENSACO350. In brief, it was subjected to thermal treatment in air at different temperatures; 530, 540 and 550 °C for 1 hour to obtain different oxidation levels. The samples thus prepared were denoted as ENSACO-Air530, ENSACO-Air540, and ENSACO-Air550, respectively. The resulting ENSACO-Air540 sample was then processed by $H_2$ annealing at 900 °C for 1 hour to remove the oxygen functional groups and was designated as ENSACO-Air540-$H_2$. Details of all carbon materials are listed in **Table S1**.

### *2.2 Characterizations*

Thermal oxidation behavior of the CNOs and ENSACO samples was performed by thermogravimetric analysis (TGA, METTLER TOLEDO TGA 851) in an air flow (100 mL $min^{-1}$) at 900 °C (heating rate of 10 °C $min^{-1}$). The porous texture of the samples was analyzed by $N_2$ adsorption/desorption measurement at –196 °C (Micromeritics ASAP 2420). The samples were outgassed under vacuum at 200 °C for 12 hours prior to measurement. The specific surface area (SSA) was calculated using the Brunauer-Emmett-Teller (BET) method in the relative pressure ($P/P_0$) range of 0.01–0.30. Pore size distributions were obtained from the $N_2$ adsorption isotherms using two-dimensional non-local density functional theory (2D-NLDFT) with a heterogeneous surface pore model for carbon materials, implemented in SAIEUS 3.0 (39, 40). In this study, BET analysis was used exclusively for SSA determination, while NLDFT was used for pore size distribution analysis. The morphology and local structure of the as-prepared carbons was examined using high-resolution transmission electron microscopy (HRTEM, JEOL ARM200F). The particle size distribution was evaluated using ImageJ software based on HR-TEM images. The crystal structure was determined using X-ray powder diffraction (XRD, Philips X'Pert MPD diffractometer) using a Cu $K\alpha_{1,2}$ doublet and a flat-plate Bragg-Brentano theta-theta geometry. The graphitic interlayer space, *d*-spacing ($d_{002}$), was calculated from the (002) diffraction peak of graphite. The crystallite size along the *c*-axis, denoted as $L_c$ (height of the stacked plane), and the size of the layer planes, denoted as $L_a$ (width of the stacked plane), were determined using the Scherrer equation (41):

$$L = \frac{K\lambda}{\beta \cos\theta} \quad (1)$$

where $K$ is a constant specific to the crystal direction and $\beta$ is the full width at half maximum (FWHM) of the diffraction peak. $L_c$ was calculated from the (002) diffraction peak with $K$ equal to 0.90, while $L_a$ was calculated from the (100) diffraction peak with $K$ equal to 1.84.

The local carbon structure, the degree of graphitization and defects in different carbons were also investigated by Raman spectroscopy using a LabRAM 300 (Horiba Jobin-Yvon) spectrometer equipped with a solid-state Laser (Compass 315M from Coherent) with an excitation wavelength of 532 nm. Nine individual spectra were acquired and averaged to ensure a representative analysis of the material structure. The ratio of the intensities of the disordered

(D) and graphitic (G) peaks, known as ($I_D/I_G$), was calculated from the ratio of the peak areas fitted with the Gaussian function, according to the literature (42, 43). The amount of carbon edge site and active surface area (ASA) were characterized by temperature programmed desorption coupled with mass spectroscopy (TPD-MS). Briefly, the samples were heated under high vacuum ($10^{-4}$ Pa) at 950 °C with a heating rate of 5 °C min$^{-1}$ and maintained at this temperature for 1 hour to ensure complete desorption of physisorbed species and decomposition of oxygen-containing surface groups. The resulting gases, mainly CO and $CO_2$, were continuously monitored by mass spectrometry to quantify the desorbed species corresponding to the different oxygen functionalities. After the sample was cooled to 300 °C, oxygen gas was introduced for a chemisorption period of 10 hours, resulting in the formation of surface oxygen complexes at the carbon edges and defect regions. Upon further heating to 950 °C, the oxygen-containing groups were decomposed. Taking into account the number of moles of desorbed gases and the area of edge carbon edge sites that chemisorbs an oxygen atom (0.083 nm$^2$), the ASA, defined as the surface area occupied by chemisorbed oxygen, could be determined (44, 45). In addition, the surface functionality was further analyzed using deuterium-labeled TPD (46). The samples were placed in a vial and dried under vacuum. Once the vial was sealed, $D_2O$ was added to the sample using a syringe to create a suspension and then heated to 60 °C. The sample was kept at this temperature for 24 hours to allow proton exchange of the surface functional groups with deuterium. The suspension was then placed on a sample holder and freeze-dried in the TPD prior to analysis. The samples were then heated to 1,600 °C (with a heating rate of 10 °C min$^{-1}$) under high vacuum (1x10$^{-5}$ Pa) and the desorbed gas was measured using a quadrupole mass spectrometer. Deconvolution of both conventional TPD and deuterium-labeled TPD profiles was performed to distinguish the contributions of the different functional groups, according to the procedure described by Ishii et al. (46, 47) and Rocha et al. (48). Further details are given in the Supporting Information.

*2.3 Electrochemical characterizations*

The investigation of electrochemical performance was carried out with two symmetric electrodes configuration in the Swagelok cell. To prepare the electrodes, the samples were mixed with polytetrafluoroethylene (PTFE) in ethanol to form a slurry. The weight ratio of carbon to binder was 9:1 and the mixture was then evaporated overnight. The obtained self-standing film was then cut to a size of 8 mm diameter and a thickness of ~200 µm, giving an average mass loading of ~12 mg cm$^{-2}$. The electrodes were then installed in the Swagelok cell and separated by a 50 mm thick cellulose separator (GF/C, Whatmann). Cyclic voltammetry (CV) was performed from 0 to 1.8 V for 21 m LiTFSI and from 0 to 0.8 V for 1 m LiTFSI (at scan rates of 1 mV s$^{-1}$). Galvanostatic charge-discharge with potential limitation (GCPL) was measured at a current density of 0.1–10 A g$^{-1}$ and the specific capacitance ($C_s$, F g$^{-1}$) was calculated using the following equation:

$$C_s = 4\frac{i\,\Delta t}{m\,\Delta V} \quad (2)$$

where $i$ is the applied current (A), $\Delta t$ is the discharge time from the GCPL profiles (s), $m$ is the total mass of the active material of both electrodes (g), and $\Delta V$ is the voltage window of the supercapacitor (V).

The areal capacitance ($C_{\mathrm{areal}}$, F cm$^{-2}$) was calculated by normalizing $C_s$ with the specific surface area using the following equation.

$$C_{\mathrm{areal}} = \frac{C_s}{\mathrm{SSA}} \quad (3)$$

## 3. Results and Discussion

*3.1 Modification of carbon structures and their physicochemical properties*

CNOs and ENSACO were selected to produce a family of various carbon materials, which were subjected to a series of thermal treatment oxidation and reduction processes. First, CNOs-1500 was synthesized by the thermal treatment of nano-diamonds (NDs) at 1500 °C under an inert atmosphere. Subsequently, CNOs-1500 was oxidized in air, and the oxidation temperature was selected based on the TGA profiles (**Figure S1**) to generate surface functional groups on the carbon edge sites. This oxidation process was performed at 460 °C, a good compromise for surface functionalization and reasonable burn-off of carbon. The result was the sample CNOs-1500-Air with a weight loss of about 24 % (**Table S1**). Subsequently, a second thermal treatment was carried out under a hydrogen atmosphere at 900 °C to reduce the functional oxygen surface group while preserving the active sites, resulting in the CNOs-1500-Air-$H_2$ sample.

A similar procedure was used with the ENSACO. However, due to the different reactivity of the two materials in an air environment caused by their different particle size, porosity and structure (see TGA profiles in **Figure S1**), several oxidation temperatures were tested, i.e., of 530 °C, 540 °C, and 550 °C, resulting in weight losses of 17.9 % for ENSACO-Air530, 32.4 % for ENSACO-Air540, and 44.5 % for ENSACO-Air550. To ensure efficient oxidation treatment while avoiding excessive mass loss, a weight loss of around 30 % was targeted for both CNOs and ENSACO carbons. ENSACO-Air540, with a similar weight loss range as CNOs-1500-Air, was selected for the subsequent second thermal treatment at 900 °C in a hydrogen atmosphere, resulting in the formation of the ENSACO-Air540-$H_2$ sample. The details of the thermal treatment and percentage weight loss of CNOs and ENSACO materials are summarized in Supporting Information (**Figure S2** and **Table S1**).

Insights into the morphology of the prepared carbon were determined by HR-TEM (see all images in **Figure S3**). The CNOs consist of the aggregation of very small nanoparticles. The CNOs particles exhibit a spherical shape formed from concentrically stacked layers of graphitic planes (**Figures 1(a)** and **1(b)**) (49). This confirms the formation of the onion-like structure, having the average size of 5–6 nm. Moreover, the average size distribution of CNOs-1500 increases slightly during oxidative thermal treatment (**Figure 1(c)**), which is due to the enrichment of surface chemistry and the number of defects at the boundaries of graphitic planes. However, the size of CNOs-1500-$H_2$ was decreased after the second thermal treatment, which reduces the presence of oxygen functional groups on the carbon surface.

In contrast, the ENSACO particles have primarily an amorphous structure with a relatively high pore volume (**Figure 1(d-e)** and **S4**) (50). **Figure 1(e)** shows that a small fraction of the particles exhibits a concentric graphitic structure, while the majority show a core with a disordered structure. In particular, it shows distorted carbon layers concentrically arranged within the particle circumference. The average size of ENSACO particles is 37–41 nm. After thermal oxidation, we notice that some of the particles consist of a few concentrically arranged interconnected layers of graphene segments. Similarly to the size trend observed for CNOs, the particle size of ENSACO increases slightly during thermal heating in air and decreases with subsequent hydrogen treatment, which can be seen in **Figure 1(f)**.

The porous textures of the as-prepared carbon materials were determined by nitrogen adsorption-desorption measurements. The ENSACO exhibits type IV isotherms, which are characteristic of mesoporous materials (**Figure 2(a)**). The increase in $N_2$ adsorbed volume in the low-pressure region indicates the presence of micropores. Indeed, the pore size distribution show the presence of micropores (less than 2 nm), with average size of ~1 nm and mesopores (2–15 nm). These mesopores were observed by HR-TEM inside the particles, and as noticed on $N_2$ adsorption/desorption isotherms the process of thermal treatment increases the proportion of the mesopore structure (**Figure 2(c)**) as indicated by the enlargement of the hysteresis loop (**Figure 2(a)**) and the increase in mesopore volume (**Figure 2(e)**). In addition, larger pores are seen between 15 and 25 nm, corresponding to inter-grain porosity, in line with the $N_2$ isotherm which steeply increases at high relative pressure ($P/P_0$ > 0.8 nm).

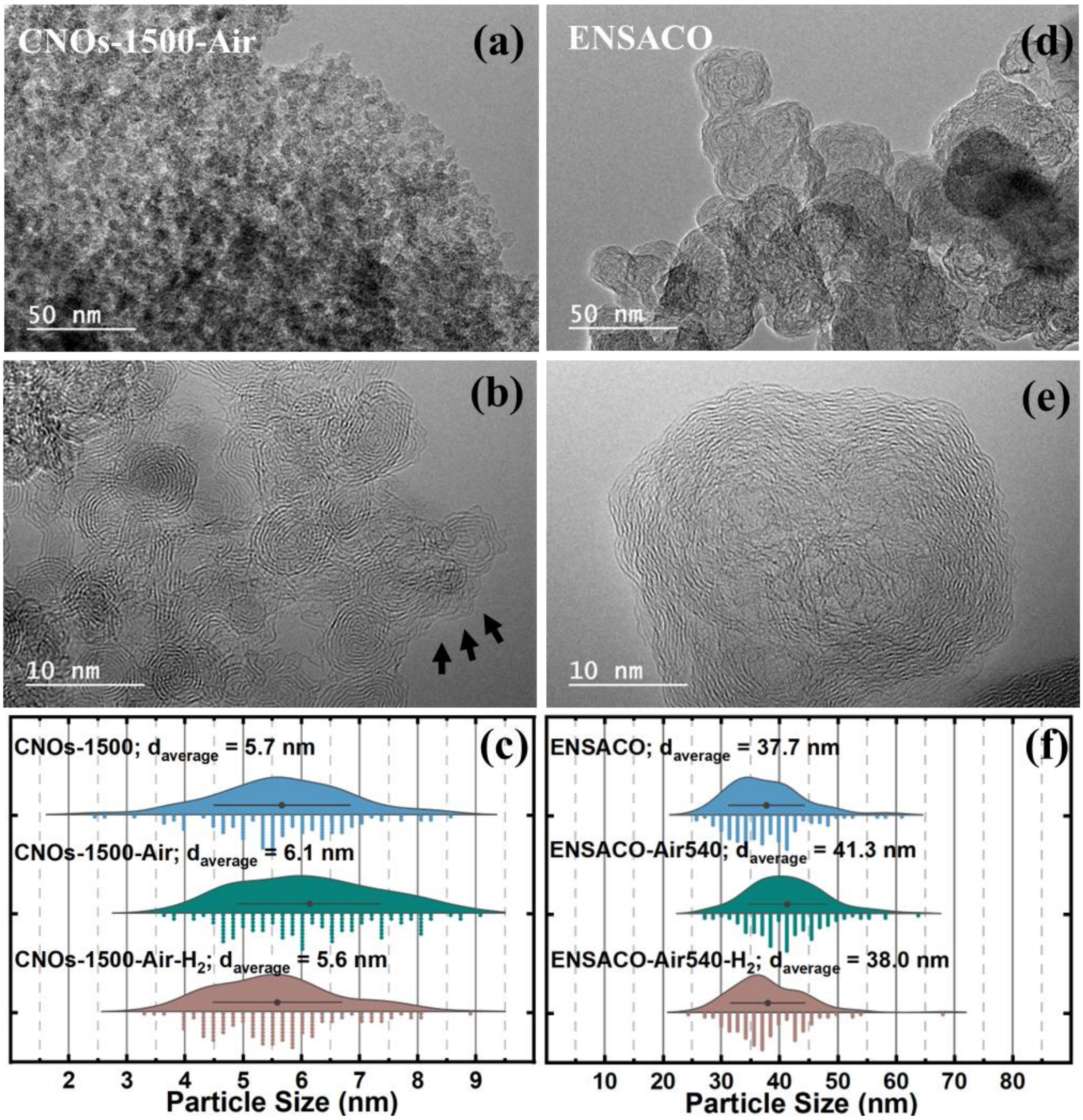

**Figure 1**. Left: HR-TEM images of (a) CNOs-1500-Air at 50 nm scale, (b) CNOs-1500-Air at 10 nm scale, (c) the particle size distribution of CNOs carbon. Right: (d) ENSACO at 50 nm scale, (e) ENSACO at 10 nm scale, and (f) the particle size distribution of ENSACO.

On the other hand, the CNOs exhibit type II isotherms with an H3 hysteresis loop, which is characteristic of non-porous or lamellar materials (**Figure 2(b)**). The pore size distribution of CNOs shows a microporous structure with a pore width of ~1 nm and mesopores of about 10 nm (**Figure 2(d)**), which correspond to the interparticle and intercluster spaces between the CNOs. This observation of such a pore distribution could agree well with HR-TEM, where the particle size of CNOs was smaller than the size of mesopores determined on CNOs structure by $N_2$ adsorption. In addition, the proportion of micro-meso pore volume are shown in **Figure 2(e)** for both ENSACO and CNOs samples after thermal treatment. For CNOs, SSA increases from 366 to 468 $m^2\ g^{-1}$ after air oxidation. This improvement is attributed to the formation of new micropores by oxidative removal of disordered carbon and the evolution of $CO_x$ gases. In contrast, ENSACO has a much higher initial SSA of 797 $m^2\ g^{-1}$, which increases to 935, 933 and 902 $m^2\ g^{-1}$ for ENSACO-Air530, Air540 and Air550 respectively. The increase in SSA after air oxidation is consistent with previous results (35, 36). On the other hand, annealing at higher temperatures may lead to excessive oxidation due to the more disordered carbon structure, resulting in an enlargement of the existing micropores or even their coalescence into mesopores. Although this transformation increases the total pore volume, it contributes less to SSA because mesopores provide less surface area per unit volume compared to micropores. These trends and mechanisms are consistent with previous studies on disordered carbons subjected to thermal treatment (51). Furthermore, annealing under $H_2$ atmosphere leads to only minor changes in SSA and pore structure for both materials **(Table 1)**, suggesting that oxidative conditions primarily determine the evolution of porosity.

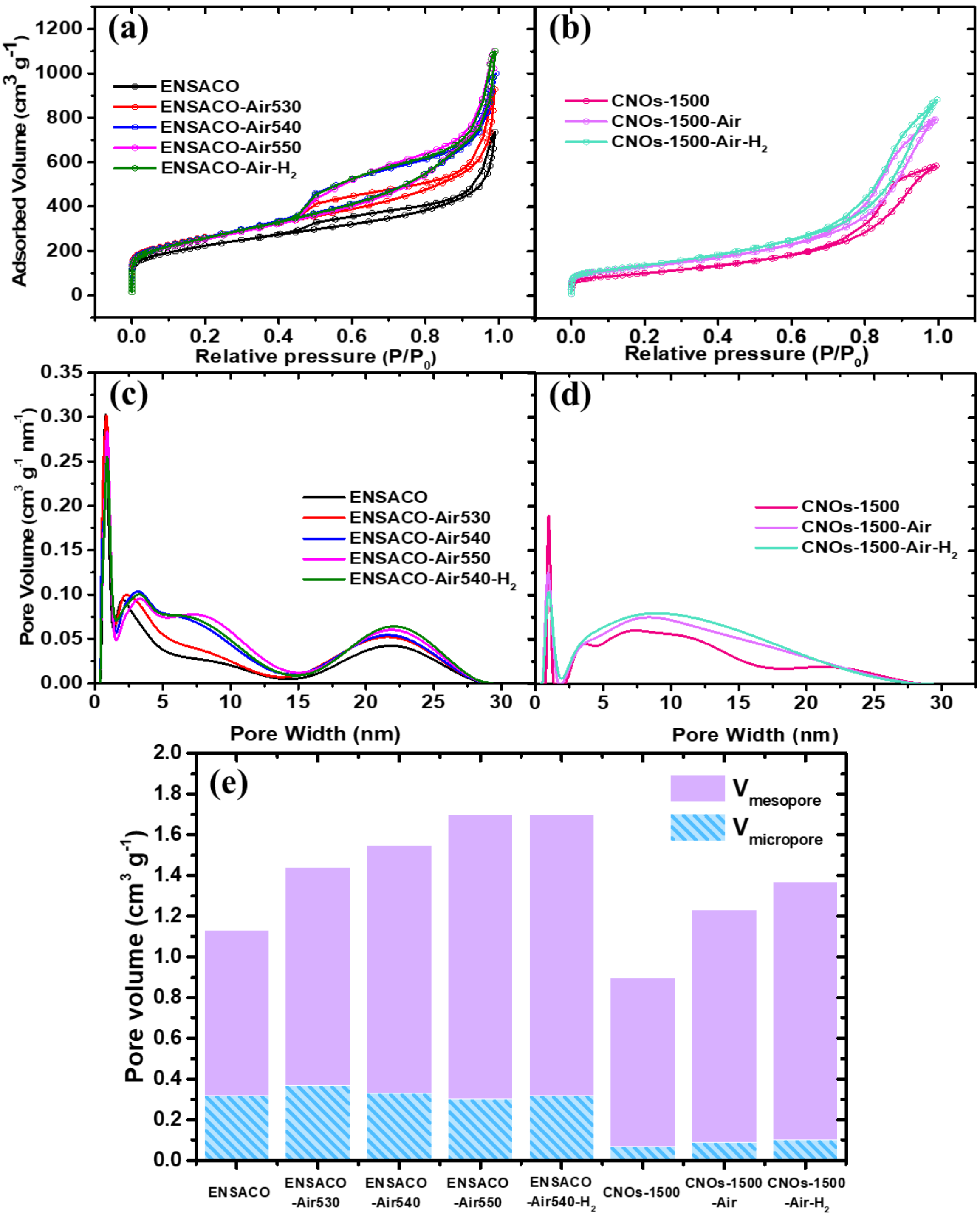

**Figure 2**. $N_2$ adsorption-desorption isotherms of (a) ENSACO and (b) CNOs samples. NLDFT pore size distribution of (c) ENSACO and (d) CNOs samples. (e) A graph bar showing the repartition of micro-meso pore volumes within the total pore volume of the as-prepared carbons.

A further investigation of the carbon structure was carried out using XRD to examine the crystallinity of the carbon samples. Their XRD patterns are shown in **Figure 3(a)**. In all samples, three main peaks are observed at approximately 26°, 44° and 75°, corresponding to the (002), (100) and (110) reflection planes of graphite, respectively (COD 9011577). The (002) peak originates from the ordered stacking of the graphene sheets, while the (100) peak comes from the in-plane diffraction of the graphene. The interlayer space ($d_{002}$) was calculated, and the values obtained are also listed in **Table 1**. CNOs have a $d_{002}$ value of 3.44–3.47 Å, which is slightly higher than $d_{002}$ value of graphite (3.354 Å). This suggests a graphitic structure, in agreement with the HR-TEM results. The broad shape of the peaks can be related to the nano-metric size of the particle, known to induce such effect. However, it can be seen that ENSACO has the highest $d_{002}$ of 3.56 Å. This is an average value of the coherent diffraction domains, but one might keep in mind that this material presents both very high-graphitic domains and more disordered-like ones (as can be seen from its HR-TEM image). The $d_{002}$ values decrease with thermal oxidation, indicating a change in the carbon structure. This effect is related to the removal of disorganized domains (Csp$^3$ hybridized and presenting higher $d_{002}$), which are more prone to oxidation and combustion (52). Thus, the graphene stacking height ($L_c$) determined from the (002) peak for CNOs-1500 and ENSACO is ~14.4 and ~16.3 Å, respectively, which corresponds to about 4–5 stacked graphene layers. After thermal treatments, similar trends were observed, $L_c$ increases after the air oxidation and slightly decreases after thermal treatment with hydrogen. When comparing the two series, $L_c$ of CNOs is smaller than that of ENSACO, which could be due to its smaller particle size. Interestingly, there are samples with the same $d_{002}$ of 3.44 Å but very different in $L_c$, ENSACO-Air540-$H_2$ ($L_c$ = 16.5 Å) and CNOs-1500-Ait-$H_2$ ($L_c$ = 14.6 Å). This difference could also be because of particle size, which is higher for ENSACO and lower for CNOs. Therefore, in the case of CNOs, the low particle size would increase the peak broadness, so the FWHM ($\beta$ in Scherrer equation), and decreases the $L_c$. The peaks of the graphitic (100) planes become more prominent in the CNOs samples, reflecting the increase of graphitic crystallites in the carbon structure. The lateral graphene size ($L_a$) determined from the (100) peak shows values between 45.4–50.5 and 77.9–81.4 Å for ENSACO and CNOs carbons, respectively. Notably, as CNOs are graphitized carbon at high annealing temperatures, the lateral graphene size ($L_a$) of CNOs is thus larger than that of ENSACO.

In Raman analysis (**Figure 3(b)**), two intense peaks are observed at approximately 1,350 and 1,590 $cm^{-1}$, corresponding to the D and G bands, respectively. In general, the D band represents the defects of $Csp^2$ structure, and the G band reflects the graphite domains. These two peaks were then deconvoluted as the area ratio of the D and G bands ($I_D/I_G$) to obtain the degree of disorder and listed in **Table 1**. The results show that the $I_D/I_G$ increases with thermal treatments, indicating the increase in disorder and/or defects in the carbon structure. Notably, CNOs samples also show the second order (2D) peaks (as 5-time intensified in **Figure 3(b)**), suggesting the presence of significant number of graphitic domains (as can be seen from HR-TEM). However, ENSACO exhibits only the broad peaks in this region because of their defective structure, which is probably more predominant than the staked graphene structures. This was also supported by the low $I_D/I_G$ and the high $L_a$ of CNOs.

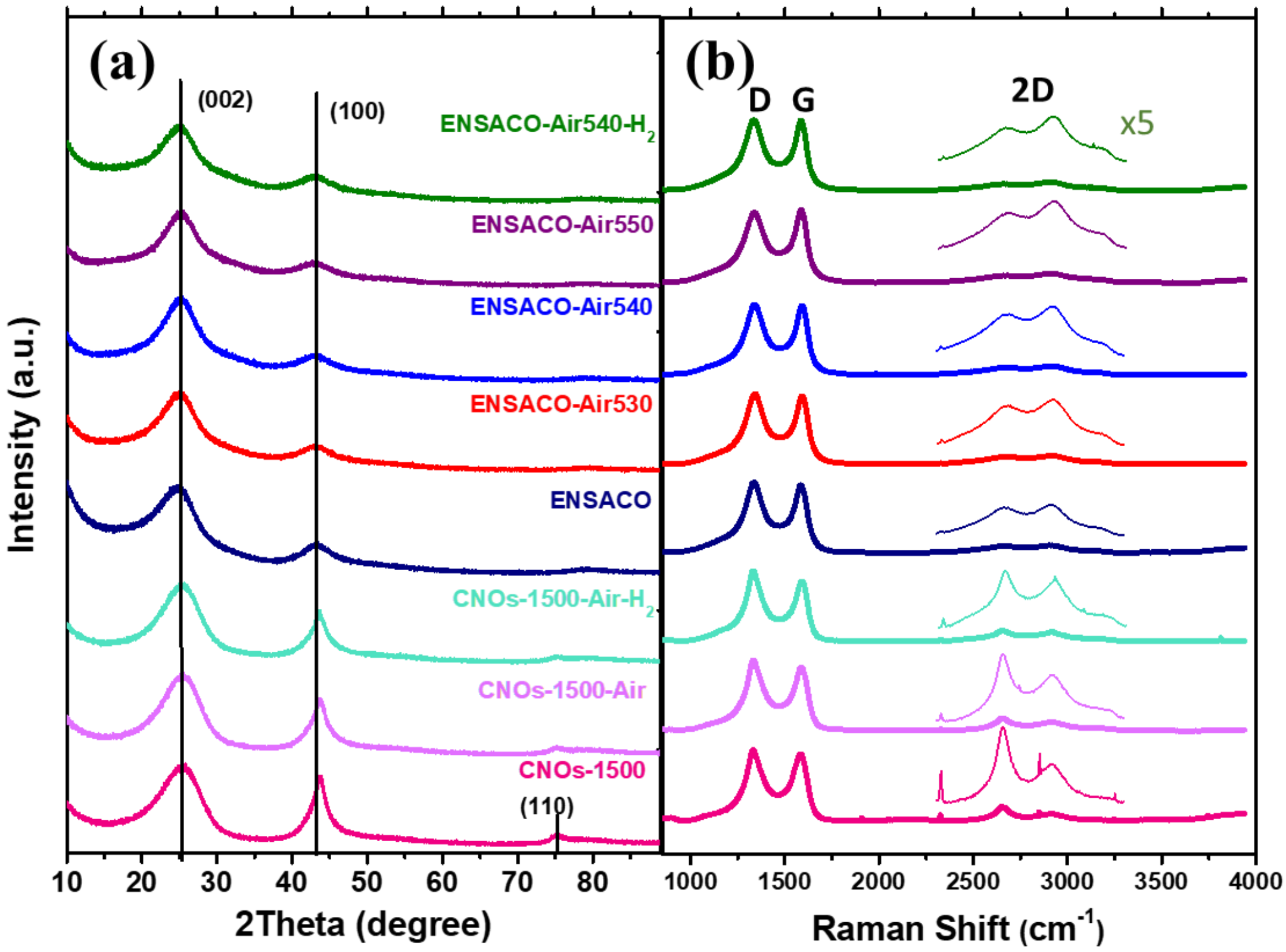

**Figure 3**. (a) XRD patterns and (b) Raman spectra of the as-prepared carbons obtained by different thermal treatments.

**Table 1**. Physicochemical and electrochemical properties of the as-prepared carbons.

| sample | SSA ($m^2$ $g^{-1}$) | ASA ($m^2$ $g^{-1}$) | DQ(CO+CO2) (mmol $g^{-1}$) | $d_{002}$ (Å) | $L_c$ (Å) | $L_a$ (Å) | $I_D/I_G$ | $C_{21m}$ LiTFSI (F $g^{-1}$) | $C_{1m}$ LiTFSI (F $g^{-1}$) |
|---|---|---|---|---|---|---|---|---|---|
| ENSACO | 797 | 15.3 | 0.22 | 3.56 | 16.3 | 47.8 | 1.86 | 38.9 | 36.0 |
| ENSACO -Air530 | 935 | 37.2 | 3.77 | 3.53 | 16.7 | 45.4 | 1.82 | 65.3 | 55.9 |
| ENSACO -Air540 | 933 | 46.1 | 4.22 | 3.51 | 16.6 | 50.5 | 1.91 | 86.4 | 70.3 |
| ENSACO-Air550 | 902 | 64.8 | 6.08 | 3.51 | 16.7 | 48.6 | 2.04 | 64.8 | 65.6 |
| ENSACO-Air540-$H_2$ | 927 | 42.3 | 0.32 | 3.44 | 16.5 | 47.2 | 1.98 | 43.6 | 33.9 |
| CNOs-1500 | 366 | 2.8 | 0.07 | 3.47 | 14.4 | 81.4 | 1.50 | 19.0 | 14.9 |
| CNOs-1500-Air | 468 | 37.9 | 1.86 | 3.47 | 14.8 | 77.9 | 1.63 | 31.0 | 28.1 |
| CNOs-1500-Air-$H_2$ | 499 | 38.7 | 0.46 | 3.44 | 14.6 | 77.9 | 1.73 | 28.8 | 22.6 |

*3.2 Analysis of carbon edge sites and surface functional groups*

Temperature programmed desorption (TPD) was used to investigate the $CO_2$ and CO desorption profiles, as well as the ASA of different carbon materials (**Figure S5**). In particular, the surface chemistry of CNOs, ENSACO, and their modified counterparts after thermal oxidation in air and thermal oxidation in air followed by hydrogen treatment was investigated. In TPD analysis performed at different temperatures (25–950 °C), $CO_2$ desorbs in a temperature range of 150 to 700 °C, while CO desorbs at higher temperatures between 400 and 950 °C. **Figure 4** reveals crucial information about the desorption gases, i.e. CO, $CO_2$, $H_2$ and $H_2O$ and ASA of all samples. It can be seen that CO-based functional groups are the predominant ones in all samples, except the $H_2$-treated ones presenting the highest $H_2$ released. ENSACO shows an increase in desorbed gases between 530 and 550 °C after thermal treatment with air, indicating enhanced surface functionality due to the modification process. Subsequent treatment with hydrogen reduces the gases released, except for $H_2$, indicating removal of surface functional groups and formation of C-H bonds. Similar trends are observed for CNOs after thermal oxidation and hydrogen treatment.

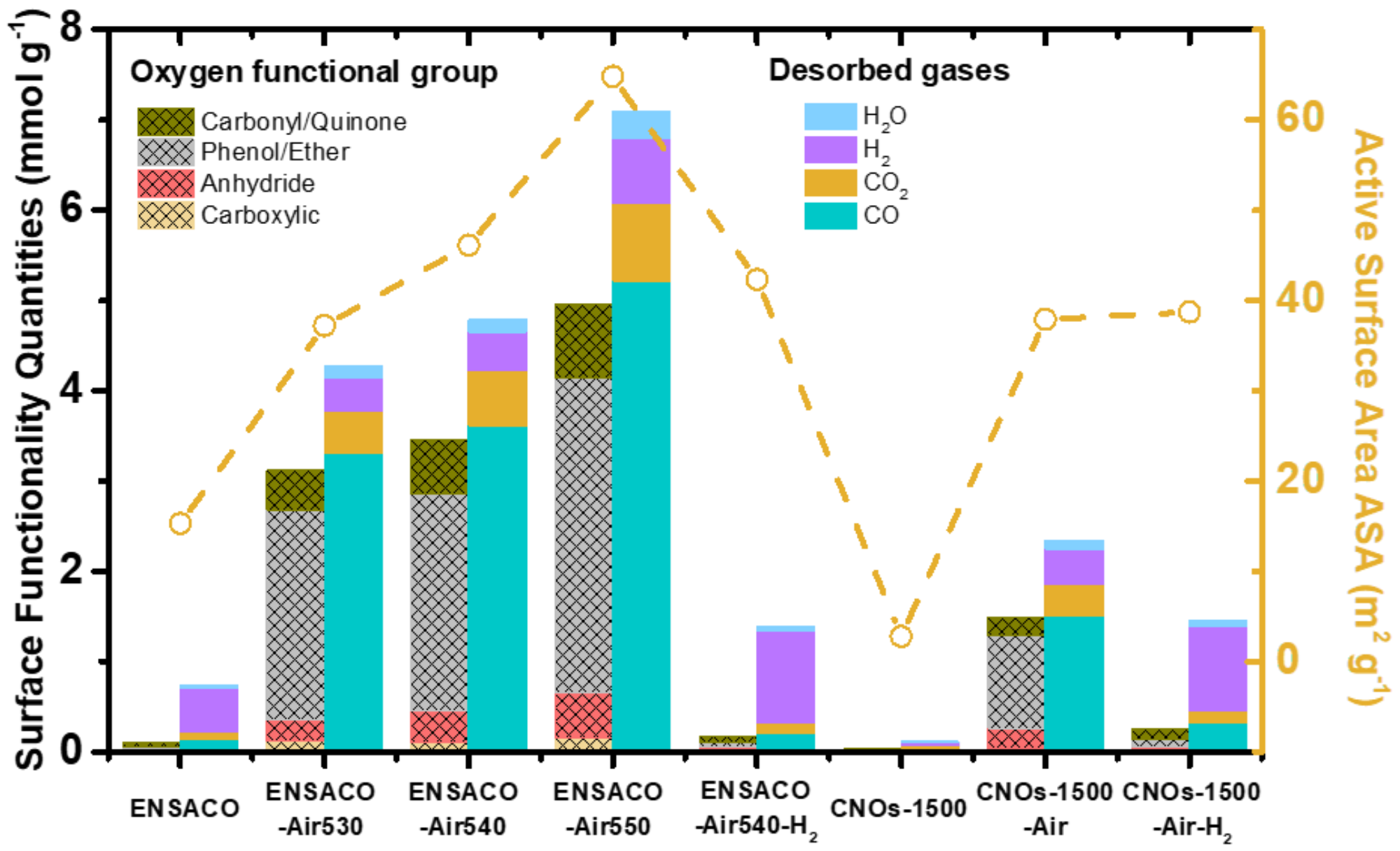

**Figure 4**. The amount of oxygen-containing functional group, desorbed gas quantities from the thermal decomposition of oxygen-containing functional groups and the ASA of various carbon samples from classical TPD.

Further analysis by ASA measurements provides information on the structural integrity and defects of carbon materials. For instance, CNOs-1500 with a low ASA value of 2.8 $m^2$ $g^{-1}$ exhibits only minimal defects, making it a reference for negligible surface functional groups and structural defects. However, after air oxidation, CNOs-1500-Air undergoes the formation of oxygen-containing functional groups and defects, which leads to a substantial increase in ASA to 37.9 $m^2$ $g^{-1}$. Surprisingly, the ASA value remains high at 38.7 $m^2$ $g^{-1}$ even after $H_2$ treatment, which removes most of the O-functional groups. This indicates that the functional groups are mainly located at defect sites, as their removal would typically lead to an increase in ASA due to defect formation (29). ENSACO, on the other hand, a carbon black obtained from the pyrolysis of hydrocarbons, initially has an ASA value of 15.3 $m^2$ $g^{-1}$, which is due to its active surface and carbon edge sites (as well as high SSA). With increasing treatment temperatures during air oxidation (ENSACO-Air530, ENSACO-Air540, and ENSACO-Air550), ASA values rise significantly (37.2, 46.1, and 64.8 $m^2$ $g^{-1}$, respectively), indicating the formation of more oxygen-containing groups and defects on the surface. Interestingly, ENSACO-Air540-$H_2$ shows no change in ASA, aligning with the observations in CNOs after $H_2$ treatment. This similarity suggests a common mechanism of how these materials respond to hydrogen treatment despite their different origins and structures.

In TPD analysis, the surface functionality of carbon material is determined by characterizing the functional groups based on their thermal decomposition behavior (53-55). It is known that ether, lactone, and anhydride decompose into CO, $CO_2$, and CO + $CO_2$, respectively. When coupled with a mass spectrometer (TPD-MS), the amount of each surface functionality can be determined by peak-fitting deconvolution, i.e. carboxyl groups decompose into $CO_2$ at 150–350 °C, phenol, and ether into CO at 500–750 °C, and anhydrides into both CO and $CO_2$ at 350-600 °C, as shown in **Figure S5**. However, distinguishing between certain functional groups, such as the phenol and the ether groups, can be difficult as both decompose to CO within a similar temperature range (46, 47). To distinguish between them, the subsequent desorption of $H_2$ must be evaluated, but this requires very high temperatures above 1,000 °C. Moreover, it is difficult to accurately determine the $H_2$ desorption at high temperature, as not only the phenol group contribute to the hydrogen sources at the edge. We therefore performed a deuterium-labeled TPD analysis in which the protic hydrogen was exchanged for deuterium, so that the phenolic groups become distinguishable by the release of $D_2$/DH after CO desorption (see Supporting Information for the detailed analysis).

Deuterium-labeled TPD desorption profiles and their corresponding deconvoluted spectra are shown in **Figures 5(a)** and **5(b)**. Based on the seven mechanisms that occur during thermal decomposition, the chemical structures at the edge sites can be classified into 12 types. **Figure 5(c)** summarizes the relationships between these 12 chemical structures and the desorbed gas species which can quantify the chemical structures within 9 relations. The amounts of desorbed gases can be used to estimate the amount of surface functionality at the edge of the carbon structure, as shown in **Figure 5(d)**. This amount of surface functionality shows the same trend as the desorbed amounts in the conventional TPD analysis. The total amount of oxygen functionality in ENSACO increases from 3.5 to 4.5 mmol $g^{-1}$ with oxidation temperature rise from 530 °C to 550 °C. The ether has the highest amount desorbed, followed by edge H-terminal, carbonyl and phenol group (Ph1). There is also a small amount of carboxyl, anhydride, and lactone groups. Subsequently, the amount of oxygen content of ENSACO-Air-$H_2$ decreases almost as much as it is present in the pristine carbon after hydrogen treatment, but the amounts of H-edge almost doubled. The presence of hydrogen at the edges allows to remove surface functionality at the edges. At the same time, after thermal oxidation in air and oxidation in air followed by hydrogen treatment, the CNOs show the same trend in oxygen functionality as in ENSACO, however, the quantities are 2 to 3 times lower.

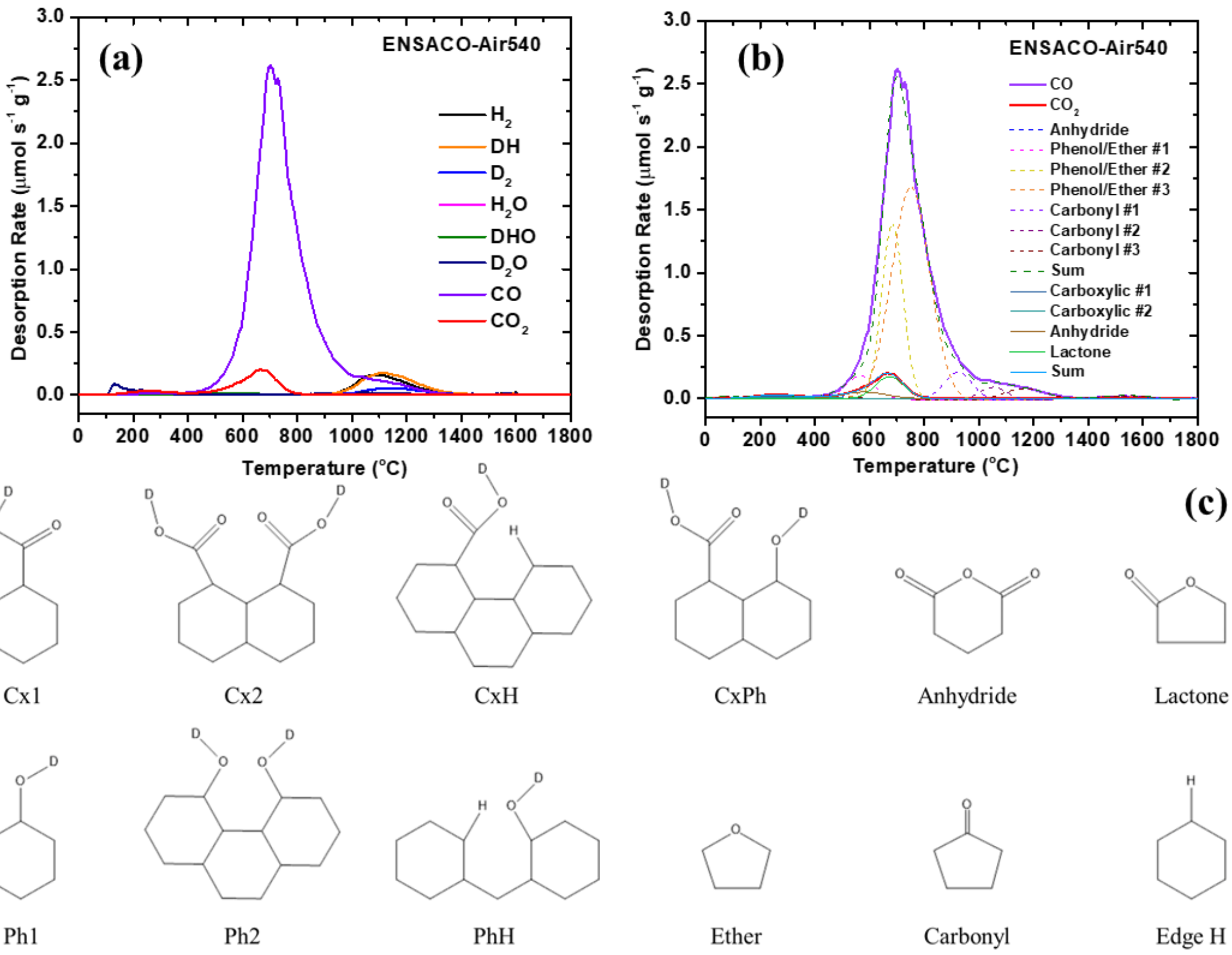

| D ($D_2$/DH) | $D_2O$ | DHO | $CO_2$ (Carboxylic) | $CO_2$ (Lactone) | CO (Anhydride) | CO (Phenol/Ether) | CO (Carbonyl) | H ($H_2$ and DH) |
|---|---|---|---|---|---|---|---|---|
| Ph1 + Cx1 | Cx2 + CxPh + Ph2 | PhH + CxH | Cx1 | CxPh + Lactone | Cx2 + Anhydride | Ph1 + Ph2 + Ether | Carbonyl + CxH | Edge H |

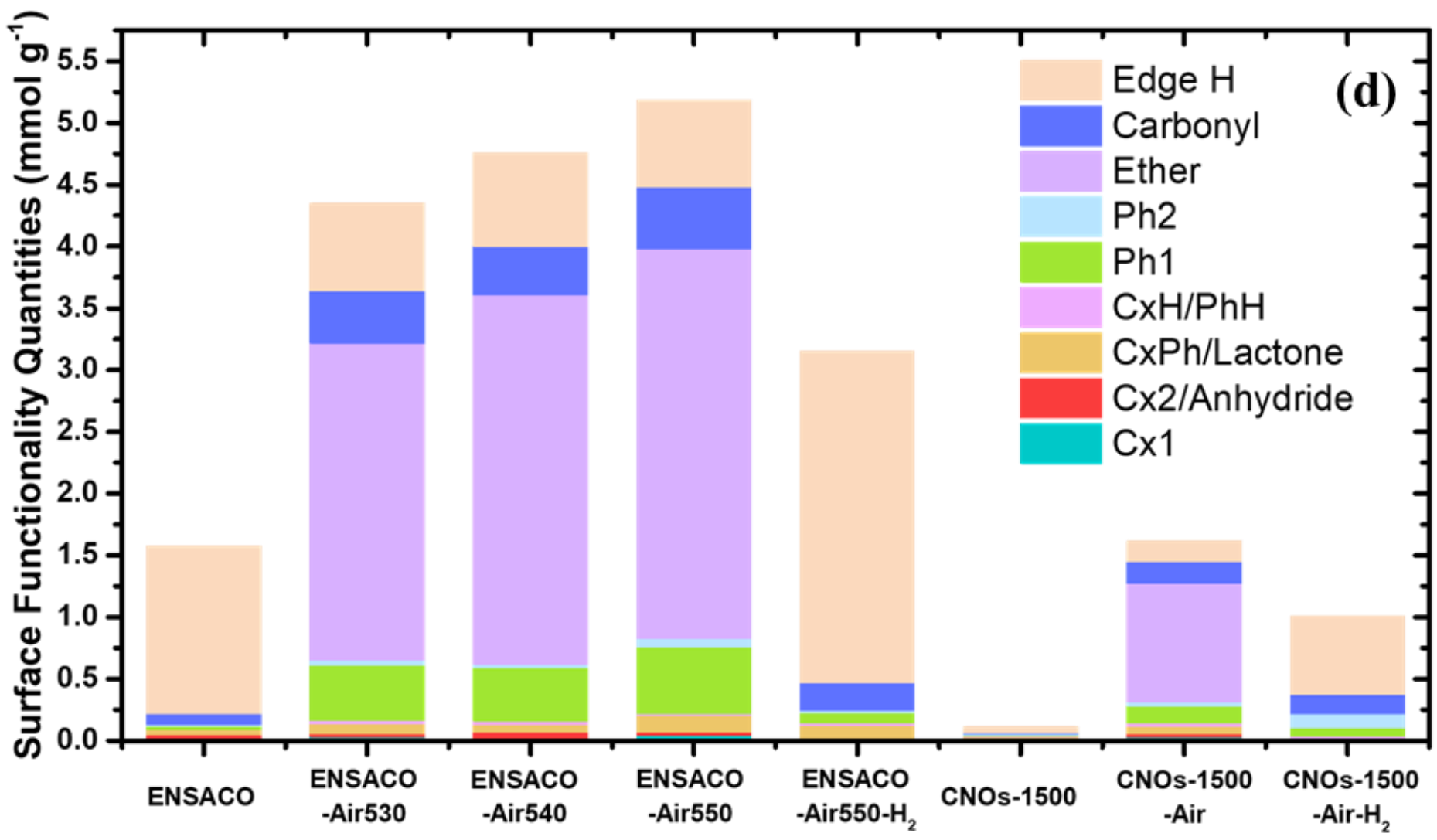

**Figure 5**. (a) An example of all gas species released from ENSACO-Air540 during the deuterium-labeled TPD and (b) deconvolution of the TPD profiles of ENSACO-Air540. (c) The expected 12 chemical structures that form at the edge of the carbon structure and their relationships between the desorbed species determined by TPD measurements and the chemical

species that form at the edge of the carbon structure. The labels in parentheses indicate the gas species from the functional source groups identified by TPD peak separation. (d) Summarized set of individual surface functionalities from the deconvolution of the TPD profiles of all materials.

*3.3 Electrochemical properties*

Electrochemical measurements of CNOs and ENSACO materials were performed using LiTFSI electrolytes (1 m and 21 m) in a symmetrical two-electrode configuration. CVs were conducted at a scan rate of 1 mV $s^{-1}$ up to various cut-off voltage values (see **Figure S7**). ENSACO-Air540 was selected as a reference for determining the maximum cut-off voltage due to its enriched surface chemistry. The 1 m LiTFSI electrolyte represented an aqueous system in which the salt serves as a solute in a water solvent. Consequently, the maximum cell voltage for CV in the 1 m LiTFSI system was limited to 0.8 V (**Figure S7(a)**) to avoid limitations associated with water decomposition at ~1.23 V. In contrast, the 21 m LiTFSI electrolyte was chosen to represent a highly concentrated LiTFSI water-in-salt system. CVs were obtained by gradually increasing the upper limit of the cell voltage from 1.4 to 2.0 V. As the voltage increases, the anodic current starts to increase at the upper limit. To mitigate undesirable reactions, the maximum cell voltage in the 21 m LiTFSI system was set at 1.8 V (**Figure S7(b)**). Notably, the CVs in both concentrations exhibit the rectangular shape, indicating EDL capacitive behaviors.

All carbons were measured with CV and GCPL to determine their electrochemical properties and compare their capacitive performance (**Figures S7(c-f)**). The CV patterns for all types of carbon have a rectangular shape, indicating a reversible process, electric double-layer capacitance (EDLC). However, at 21 m LiTFSI (**Figure S7(d)**), some samples show additional current responses near 0.8 V, indicating the presence of faradaic processes likely due to surface functional groups or defect sites. The CV profiles of ENSACO show a wider range of specific capacitance compared to the CNOs, which is due to differences in specific surface area and porous structure, which are more developed for the former one. In addition, the surface functionality of the carbon can influence the charge storage behavior by introducing redox-active sites that contribute to additional charge storage beyond pure EDLC.  This effect is evident with ENSACO-Air530, ENSACO-Air540, and ENSACO-Air550 when subjected to 21 m LiTFSI, but less observed with 1 m LiTFSI. In the 1 m LiTFSI electrolyte, the electrochemical stability window is limited by water splitting due to the dilute aqueous environment. Within this limited window, charge storage is predominantly governed by the EDLC, which does not promote the faradaic redox reactions necessary for pseudocapacitance. Conversely, the highly concentrated water-in-salt electrolyte LiTFSI, with its altered solvation structure and reduced free water activity, is expected to increase the electrochemical accessibility of surface functional groups, allowing them to contribute more effectively to surface-controlled charge storage than in the dilute 1 m LiTFSI system. This process, which enables pseudocapacitive behavior, corresponds to the strict definition of pseudocapacitive behavior as described by Brousse et al (56). Interestingly, ENSACO-Air540-$H_2$ presents similar CV as the pristine ENSACO, suggesting that the removal of O-groups negatively impacted the capacitive performance.

To further clarify the pseudocapacitive charge storage behavior, Dunn analysis was performed using CV data collected at multiple scan rates **(Figure S8)**. ENSACO-Air540 was selected as a representative sample because it clearly exhibits pseudocapacitive shapes among the ENSACO series, making it suitable for capturing general charge storage characteristics. The CV profiles maintain quasi-rectangular shapes across all scan rates in both electrolytes **(Figure S8(a))**, while log(i)–log(v) analysis at 0.6 V yields b-values close to unity (~0.90 in 1 m LiTFSI and ~0.95 in 21 m LiTFSI), indicating predominantly surface-controlled charge storage **(Figure S8(b))**. Quantitative separation of the current response shows that the surface-controlled capacitive contribution dominates over the diffusion-controlled term, with a more pronounced surface contribution in 21 m LiTFSI **(Figure S8(c))**. The absence of distinct redox peaks or voltage plateaus further indicates that the enhanced charge storage does not involve battery-type diffusion-controlled faradaic reactions. The Dunn analysis distinguishes surface-confined from diffusion-controlled processes but does not identify specific redox reactions. The enhanced surface-controlled contribution observed in the water-in-salt electrolyte, together with the presence of oxygen-containing surface functional groups identified by TPD analysis and the reduced capacitance after hydrogen treatment, supports attributing this behavior to fast, reversible surface-confined faradaic interactions associated with surface functionalities, consistent with pseudocapacitive charge storage coexisting with electric double-layer capacitance.

The galvanostatic charge-discharge plots for both CNOs and ENSACO at a constant current density of 0.1 A $g^{-1}$ are also shown in **Figures S7(e)** and **S7(f)**. It can be observed that the discharge time of CNOs increases after the first thermal treatment in air, as the comparison between CNOs-1500 and CNOs-1500-Air shows. In addition, the discharge time decreases after the second thermal treatment with $H_2$, as observed with CNOs-1500-$H_2$. The difference in discharge time indicates the cell capacitance of each CNOs state, where CNOs-1500-Air was able to increase the capacitance, influenced by the change in surface structure, chemistry, and number of defects. A similar behavior was also observed for the ENSACO materials (**Figures S7(e)** and **S7(f)**).

### *3.4 Dependence correlation of carbon structures, defects, and surface functionality on capacitance*

The capacitive performance of supercapacitors generally depends on various parameters of the electrode materials, such as crystallinity, pore structure, surface area, and surface functionality (33, 57-61). We thus first investigated the effects of crystallinity on capacitance by correlating $L_c$, $L_a$, and $I_D/I_G$ with specific capacitance ($C_s$). The correlations presented here are intended to highlight general positive or negative trends rather than establish strict linear relationships, since the crystallinity of carbon materials is intrinsically linked to other structural features – such as defect density, surface accessibility, and active surface area – that also influence $C_s$. As a result, noticeable scatter in the data is expected, especially when different carbon architectures (CNOs and ENSACO) are considered together. In **Figures 6(a)** and **6(b)**, it can be seen that the larger $L_c$, the increase in $C_s$ for both 1 m and 21 m LiTFSI. The increase of $C_s$ with an increase in $L_c$ (for instance, number stacked graphene layers) can be ascribed to the increase in quantum capacitance, attributing to the increase of the density of states when more graphene layers are added (62). However, it is also possible that the increased capacitance at higher $L_c$ is influenced by the presence of structural defects, not only at the graphite edges but also on the basal planes (as shown by the increase of ASA, SSA, and $I_D/I_G$ ratio with $L_c$). These defects may introduce additional faradaic contributions, potentially leading to pseudocapacitive behavior. Such defect-induced redox reactions may enhance charge storage beyond the electrical double layer capacitance (EDLC). The larger $L_a$ indicates an increase in the lateral size of the graphitic sheets and $C_s$ decreases with $L_a$ for both 1 m and 21 m LiTFSI (**Figures 6(c)** and **6(d)**). Therefore, less crystalline materials with more defects enhance the $C_s$. This can be further supported by the correlation between $C_s$ and $I_D/I_G$ in 1 m and 21 m LiTFSI electrolytes (**Figures 6(e)** and **6(f)**). As the degree of disorder (or $I_D/I_G$) increases, the $C_s$ also increases. These results are consistent with those reported in the literature, indicating that more disordered carbons with smaller graphene-like domains exhibit higher capacitances due to more efficient ion storage in the pores (63-66). Unlike these cited works dedicated to activated carbons, here graphitic carbons demonstrate similar behavior between $C_s$ and $I_D/I_G$, which was not previously shown.

**Figure 7** examines the influence of surface-related properties on specific capacitance ($C_s$) by comparing oxidized and non-oxidized carbon materials. This distinction enables separate evaluation of the impact of oxidation-induced surface functionalities and defect-related active sites from the intrinsic carbon framework. The correlations between $C_s$ and edge hydrogen content, active surface area (ASA), and specific surface area (SSA) are presented for both electrolyte concentrations. Linear regressions are shown to assess whether each parameter tends to increase or decrease with $C_s$, and the resulting $R^2$ values indicate varying correlation strengths. When $R^2$ values are relatively low (e.g., $R^2 < 0.5$), the lack of a precise linear relationship suggests that the parameter under consideration is not the sole factor governing capacitance, reflecting the multivariate nature of charge storage in carbon electrodes. Nevertheless, these correlations are useful for identifying general positive or negative trends and for comparing the relative influence of surface-related parameters between oxidized and non-oxidized samples.

Interestingly, $C_s$ exhibits a clear positive correlation with edge H, particularly in the oxidized samples, further highlighting the role of active sites on carbon edge planes in enhancing capacitance (**Figures 7(a)** and **7(b)**), a finding that has not been reported until now. The influence of defects in the form of ASA on $C_s$ is also examined in **Figures 7(c)** and **7(d)**. $C_s$ tends to increase with ASA for both concentrations, where ASA represents the reactivity of carbon edge plane, allowing the improvement of interactions with the electrolyte and in consequence the capacitance behavior (29).

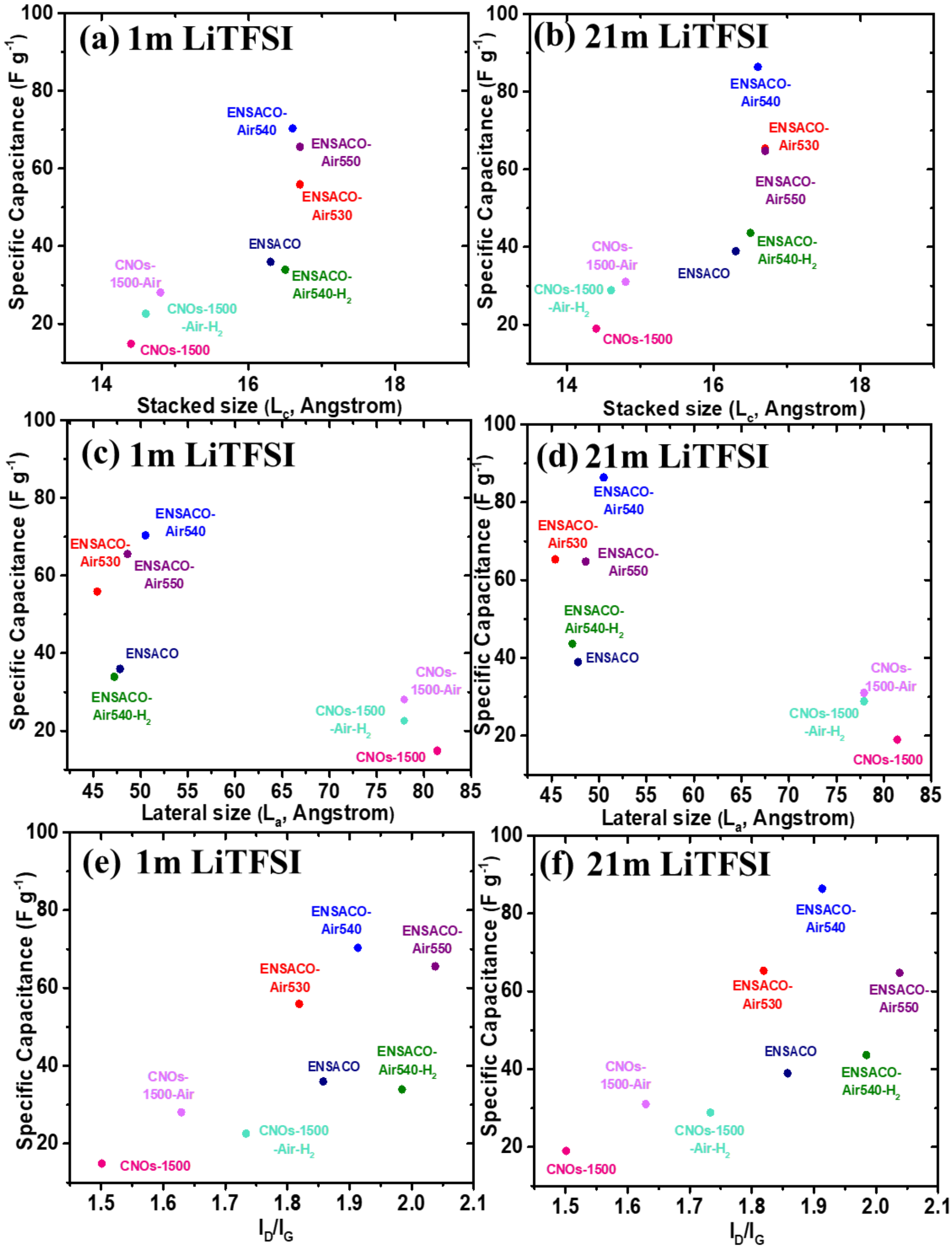

**Figure 6**. Correlation of specific capacitance ($C_s$) of the as-prepared carbon against the following parameters in 1 m and 21 m LiTFSI: (a) and (b) stacked graphene size ($L_c$), (c) and (d) lateral graphene size ($L_a$), (d) and (e) disorder degree ($I_D/I_G$).

Noteworthy, it is still difficult to see a clear correlation as there are many types of active sites on the materials (67). Nonetheless, all information regarding the defects obtained from XRD, Raman, and TPD analyses leads to the same conclusion: defects increase $C_s$. In addition, **Figures 7(e)** and **7(f)** show a positive correlation between $C_s$ and SSA for both LiTFSI electrolyte concentrations, which is common for carbon-based supercapacitors as $C_s$ is proportional to SSA not only in the typical organic electrolytes (63, 68-70), but also in this system, based on the following equation:

$$C_s = \frac{\varepsilon_r \varepsilon_0}{d} A \quad (4)$$

where $\varepsilon_r$ is the electrolyte dielectric constant, $\varepsilon_0$ is the permittivity of a vacuum, $A$ is the electrode specific surface area, and $d$ is the effective thickness of the electric double layer (71). Additionally, the oxidized samples show higher $C_s$ than the non-oxidized ones, owing to the higher SSA. Nonetheless, as SSA comprises both ASA and surface of basal plans, both contribute to $C_s$. We also investigated capacitive performance in different pore regions (**Figures S9**). Between 0.95 and 1.1 nm, $C_s$ gradually increases with pore size, while below 0.95 nm, $C_s$ shows a sharp rise with decreasing pore size, consistent with findings reported in the literature (71).

We thus investigated the effect of oxygen-containing functional groups (quantified by desorbed quantity of $CO$+$CO_2$) on the capacitance (**Figures 8(a)** and **8(b)**). As expected, $C_s$ increases with the desorbed quantity. This can be explained by the following two reasons. Firstly, the oxygen-containing functional groups on the surface of carbon materials can adsorb the solvent, and a subsequent desolvation process occurs, which allows ions to rapidly enter the unused pores, thereby increasing the capacitance (72, 73). Secondly, several mechanisms have been proposed to be responsible for the enhancement of capacitance values based on the introduction of redox pseudocapacitance (10, 15, 74). Similar trends can be observed when normalizing $C_s$ with SSA ($C_{areal}$), as shown in **Figures S10(a)** and **S10(b)**.

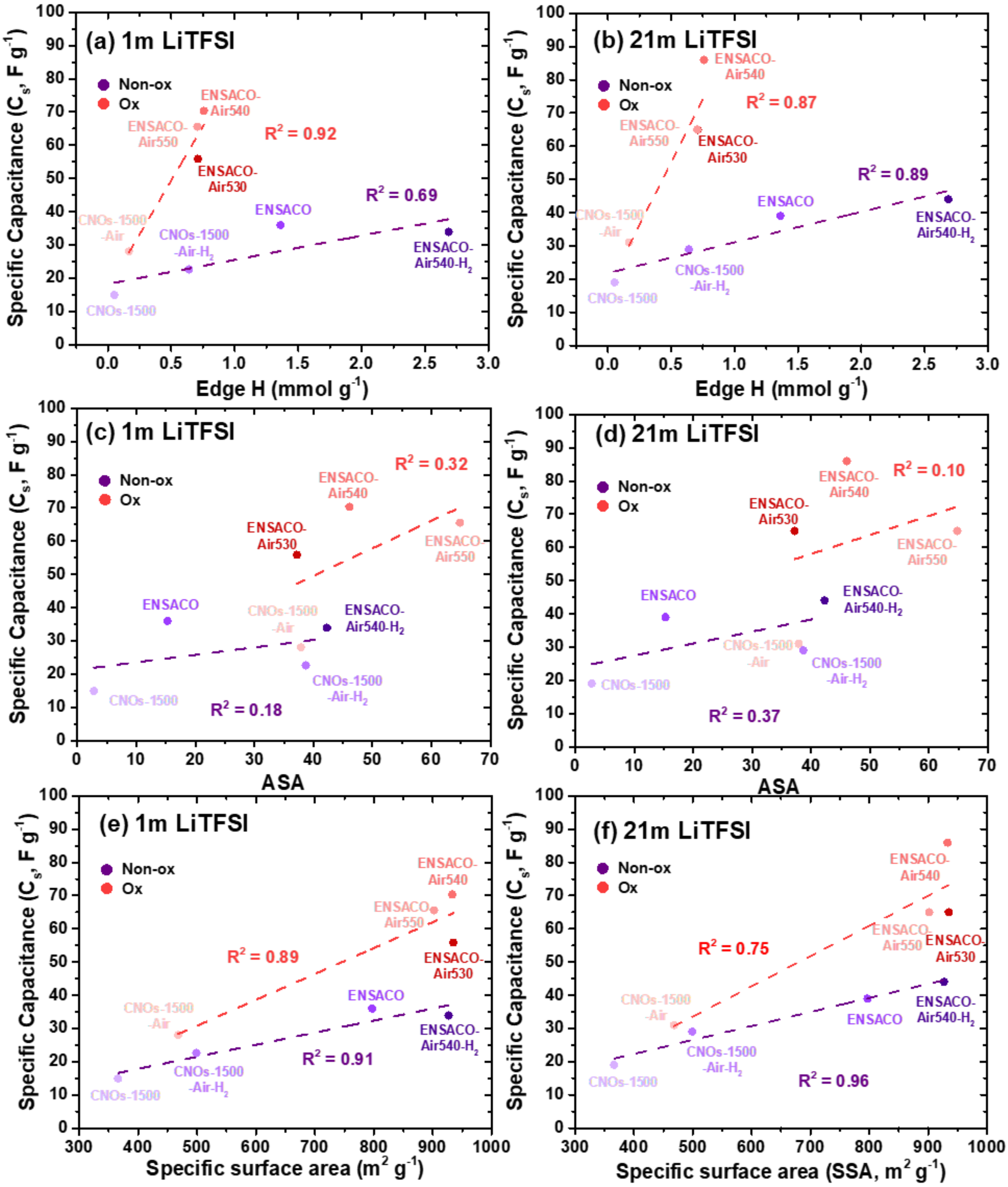

**Figure 7**. The plots of specific capacitance ($C_s$) against the following parameters in 1 m and 21 m LiTFSI: (a) and (b) the edge hydrogen, (c) and (d) ASA, (e) and (f) SSA. The samples were categorized into two series: non-oxidized (orange) and oxidized samples (purple).

The 12 chemical structures identified from the deuterium-labeled TPD were then grouped into six different surface functionalities, as shown in **Figure 5(c)**: 1) carboxylic groups are quantified by the relation of Cx1; 2) anhydrides by the relation of Cx2 + anhydride; 3) carbonyl groups by the relation of CxH + carbonyl; 4) phenols by the relation of Ph1 + Ph2; 5) ethers by the relation of ether; and 6) lactones by the relation of CxPh + lactone. This classification allows a clearer exploration of the relationship between each surface functionality and the capacitance. The contribution of oxygen to pseudo-Faradaic effects has already been studied in the literature where the oxygen-containing functional groups have a profound effect on the increase of the energy storage of supercapacitors with strongly acidic and alkaline electrolytes (75). It has been reported that alkaline functional groups (such as carbonyl and quinone groups) play a particularly significant role, reacting with protons ($H^+$) in acidic aqueous solutions (76-79), while acidic functional groups (such as phenol and carboxyl groups) react with hydroxyl ions in alkaline aqueous solutions (80, 81). Although it was reported that the oxygen-containing functional groups have a limited contribution to the capacitive performance with neutral electrolytes (75), our results clearly shows a positive impact of oxygen functionality on the capacitance in the LiTFSI electrolytes, having the pH of 6.4-7.3. **Figures 8(c)** and **8(d)** show a relationship between surface functionality and $\Delta C_s$, which is the differences between $C_s$ of the oxidized and non-oxidized samples. Interestingly, the relationship between surface functionality and $\Delta C_s$ can be categorized into 3 groups (except lactone, whose relation is unclear): (1) carboxylic and anhydride groups, (2) carbonyl and phenol groups, and (3) ether.

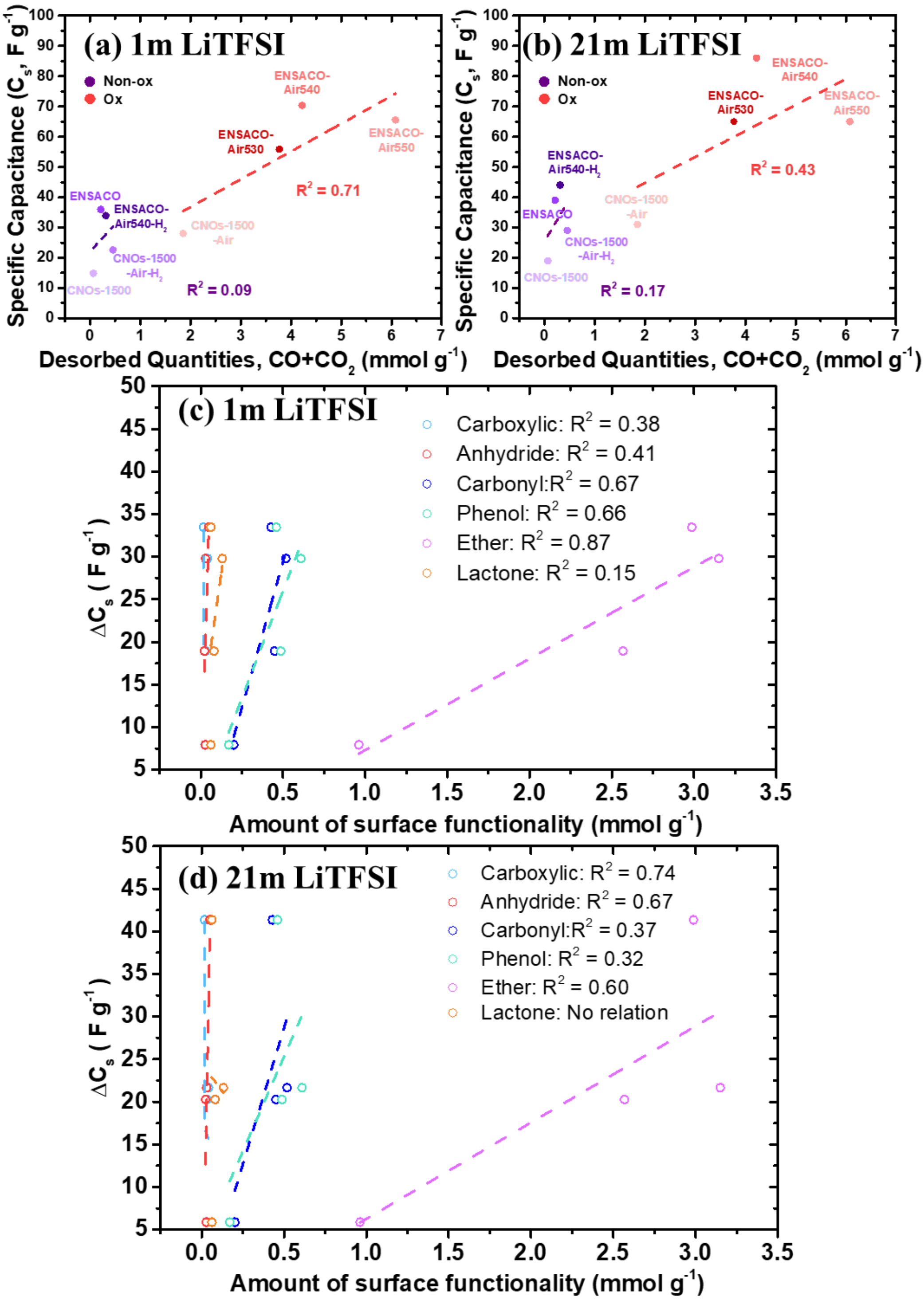

**Figure 8**. Specific capacitance ($C_s$) against desorbed quantities of $CO+CO_2$ in (a) 1m and (b) 21m LiTFSI and the correlation of $\Delta C_s$ with the amount of surface functionality in (c) 1 m LiTFSI and (d) 21 m LiTFSI.

A drastic increase of $\Delta C_s$ is observed in **Figures 8(c-d)** with carboxylic and anhydride groups. Our results pointed out different effects from carbon structures and defects on supercapacitive performance in LiTFSI electrolytes. Specific capacitance ($C_s$) increases with stacked size ($L_c$), which can be ascribed to the increase in quantum capaciatnce. However, carbon with less crystallinity (low lateral size, $L_a$, and high degree of disorder $I_D/I_G$) leads to the increase in $C_s$. Both SSA and ASA have the positive impact on $C_s$, as $C_s$ is generally proportional to SSA and ASA represents the reactivity of carbon edge plane, allowing the improvement in ion adsorption. This can be further supported by the postive influnce of edge H on $C_s$, emphasizing the role of active sites of carbon edge planes in enhancing the capacitance. In addition, our results shed light on a prominent surface functional group that affects supercapacitive performance in LiTFSI electrolytes. Although $C_s$ increases with the amount of oxygen-containg functional groups (in terms of CO + $CO_2$ desorbed quantities), we found that carboxylic and anhydride groups have the great impact in increasing the capacitance. It was reported that the solavated $Li^+$ can be completely desolvated when the pore size is less than 3.61 Å (72). However, both CNOs and ENSACO samples have the pore of larger than 3.61 Å (**Figures 2(c-d)**). Therefore, such desolvation effects will not be taken into account in our case. In addition, the solvation size of $Li^+$ ions is still smaller than the pore of CNOs and ENSACO (82). We thus suggest that solvated $Li^+$ ions can readily access the pore structure and interact with surface functional groups, particularly carboxylic and anhydride groups, which are predominantly located at carbon edge and defect sites. Our findings thus led to the optimization of not only the carbon structure but also the surface of the carbon electrode to enhance supercapacitor performance.

## 4. Conclusion

This research highlighted the critical role of graphitic nanocarbon structural features, defects, and surface functional groups in determining the electrochemical performance of supercapacitors in LiTFSI electrolytes, which has often gone unnoticed so far. Our findings indicated that the performance is not only determined SSA and ASA, but also by a complex interplay of chemical and structural properties. In particular, specific oxygen functionalities, especially carboxylic acid and anhydride groups, strongly affect ion interactions at the electrode–electrolyte interface. In addition, crystallinity and graphitic characteristics also influence capacitance, with smaller graphene domains and higher disorder favoring ion storage. The trends observed across different carbon materials show that the targeted modification of the nanocarbon architecture can compensate for the surface area limitations, allowing a refined strategy for designing functional electrodes. Compared to WiSE systems reported in the literature, our findings provide a deeper insight into subtle structural modifications that lead to measurable changes in electrochemical behavior - an aspect that remains under-researched. These results underscore the importance of tailoring carbon electrode properties to optimize supercapacitor performance. The obtained property-capacitance relationships provide valuable guidance for the development of advanced electrode materials in next-generation energy storage devices.

## Acknowledgements

This research was supported by Thailand Science Research and Innovation (TSRI) Fundamental Fund, fiscal year 2024, and has received funding support from the NSRF via the Program Management Unit for Human Resources & Institutional Development, Research and Innovation (Contract No. B13F660086). P.S acknowledges the support from Thammasat Post Doctoral Fellowship (Contract No. TUPD 4/2567). The authors acknowledge the laboratory of excellence for electrochemical energy storage STORE-EX LabEX (ANR-10-LABX-76-01) and the people from IS2M platforms for their experimental help: Jean-Marc Le Meins (XRD), Cyril Vaulot (Adsorption), Loic Vidal (HRTEM), and Simon Gree (Raman).

## Author contributions

Phongphot Sakulaue: Methodology, Investigation, Data Curation, Writing - Original Draft, Visualization. Bénédicte Réty: Investigation, Data Curation. Takafumi Ishii: Investigation, Data Curation. Camelia Matei Ghimbeu: Conceptualization, Methodology, Validation, Formal analysis, Writing - Review & Editing, Supervision, Project administration, Funding acquisition. Khanin Nueangnoraj: Conceptualization, Methodology, Validation, Formal analysis, Writing - Review & Editing, Supervision, Project administration, Funding acquisition.

## Conflicts of interest

There are no conflicts to declare.

## Data availability

The data supporting this article have been included as part of the Supplementary Information.